\documentclass[sigconf]{acmart}
\AtBeginDocument{%
  }

\copyrightyear{}
\acmYear{}
\acmDOI{}

\acmConference{}
\acmPrice{}
\acmISBN{}
\setcopyright{none}

\usepackage{listings}
\usepackage{xcolor}
\usepackage{framed}
\usepackage{amsmath}
\usepackage{enumitem}
\usepackage{caption}
\usepackage{float}

\definecolor{keywordcolor}{rgb}{0.7, 0.1, 0.1}   
\definecolor{stringcolor}{rgb}{0.2, 0.5, 0.2}    
\definecolor{commentcolor}{rgb}{0.4, 0.4, 0.8}   
\definecolor{backcolor}{rgb}{0.98, 0.98, 0.98}   

\lstdefinestyle{JSONstyle}{
  backgroundcolor=\color{backcolor},
  basicstyle=\ttfamily\footnotesize,
  breakatwhitespace=false,
  breaklines=true,
  captionpos=b,
  commentstyle=\color{commentcolor},
  keepspaces=true,
  keywordstyle=\color{keywordcolor}\bfseries,
  language=Java,
  morekeywords={true, false, null},
  numbersep=10pt,
  numberstyle=\tiny\color{gray},
  rulecolor=\color{black},
  showspaces=false,
  showstringspaces=false,
  showtabs=false,
  stringstyle=\color{stringcolor},
  tabsize=2,
  frame=single,
  frameround=tttt,
  framesep=5pt,
  xleftmargin=15pt,
  framexleftmargin=10pt,
}

\lstdefinestyle{promptStyle}{
  basicstyle=\ttfamily\small,
  backgroundcolor=\color{gray!10},
  frame=single,
  framerule=0.4pt,
  rulecolor=\color{gray!50},
  breaklines=true,
  breakatwhitespace=true,
  captionpos=b,
  numbers=none,
  xleftmargin=5pt,
  xrightmargin=5pt
}

\lstnewenvironment{promptlisting}[2][]
  {\lstset{style=promptStyle,title={\textbf{\textcolor{brown}{#2}}},#1}}
  {}

\newfloat{kcfloat}{tbhp}{kc}
\floatname{kcfloat}{Listing}

\newenvironment{graybox}[2]{%
  \begin{kcfloat}
  \begin{framed}
  \noindent\colorbox{gray!20}{%
    \begin{minipage}{\dimexpr\linewidth-2\fboxsep\relax}
    \textbf{\color{brown}#1}\hfill\textbf{#2}
    \end{minipage}%
  }\\[0.5em]
}{%
  \end{framed}
  \end{kcfloat}
}

\usepackage{mdframed}

\newenvironment{simplebox}[1][]{%
  \begin{mdframed}[
    linewidth=0.5pt,
    linecolor=black,
    backgroundcolor=white,
    innerleftmargin=10pt,
    innerrightmargin=10pt,
    innertopmargin=5pt,
    innerbottommargin=5pt,
    skipabove=5pt,
    #1
  ]
}{%
  \end{mdframed}
}

\renewcommand\footnotetextcopyrightpermission[1]{}
\begin{document}

\title{Mining the Gold: Student-AI Chat Logs as Rich Sources for
Automated Knowledge Gap Detection}
\author{Quanzhi Fu}
\affiliation{%
  \institution{Virginia Tech}
  \city{Blacksburg}
  \state{VA}
  \country{USA}
}
  \email{quanzhif@vt.edu}

  \author{Qiyu Wu}
\affiliation{%
  \institution{Independent Researcher}
  \city{Dallas}
  \state{TX}
  \country{USA}
}
  \email{qiyu.wu71@gmail.com}

\author{Dan Williams}
\affiliation{%
  \institution{Virginia Tech}
  \city{Blacksburg}
  \state{VA}
  \country{USA}
}
\email{djwillia@vt.edu}


\begin{abstract}
With the significant increase in enrollment in computing-related programs over the past 20 years, lecture sizes have grown correspondingly. In large lectures, instructors face challenges on identifying students' knowledge gaps timely, which is critical for effective teaching. Existing classroom response systems rely on instructor-initiated interactions, which limits their ability to capture the spontaneous knowledge gaps that naturally emerge during lectures. With the widespread adoption of LLMs among students, we recognize these student-AI dialogues as a valuable, student-centered data source for identifying knowledge gaps. In this idea paper, we propose QueryQuilt, a multi-agent LLM framework that automatically detects common knowledge gaps in large-scale lectures by analyzing students' chat logs with AI assistants. QueryQuilt consists of two key components: (1) a Dialogue Agent that responds to student questions while employing probing questions to reveal underlying knowledge gaps, and (2) a Knowledge Gap Identification Agent that systematically analyzes these dialogues to identify knowledge gaps across the student population. By generating frequency distributions of identified gaps, instructors can gain comprehensive insights into class-wide understanding. Our evaluation demonstrates promising results, with QueryQuilt achieving 100\% accuracy in identifying knowledge gaps among simulated students and 95\% completeness when tested on real student-AI dialogue data. These initial findings indicate the system's potential for facilitate teaching in authentic learning environments. We plan to deploy QueryQuilt in actual classroom settings for comprehensive evaluation, measuring its detection accuracy and impact on instruction. 
\end{abstract}




\maketitle
\section{Introduction}

Timely identification of student knowledge gaps is crucial for effective teaching~\cite{henderson2019conditions} in college education. As enrollment in computing-related programs has increased significantly in the past 20 years, class sizes have grown correspondingly larges~\cite{batten2025infographic}. In 2019, UC Berkeley reported that the enrollment of their introductory computer science course reached 2000 students~\cite{chen2019enrollment}. Traditional classroom approaches like in-class interactions or after-class assessments, both face significant limitations for lectures at this scale~\cite{ouhbi2020software,offerdahl2014student}. This makes instructors in computer science struggle to maintain awareness of students' common knowledge gaps and impact the teaching quality.

To address this challenge, people developed student response systems (SRS) ~\cite{polleverywhere2024platform, vevox2024classroom, socrative2024platform, macmillan2024iclicker} which help instructors conduct timely evaluation through lightweight in-class activities. However, their effectiveness is highly depends on the instructor experience to design high-quality questions and integrate them into lectures seamlessly~\cite{beatty2006designing} and improper usage of such system could actualy impede the learning outcome.~\cite{nielsen2013teaching, wood2020systematic}

Studies have shown that over 70\% of university students now use LLM-based tutoring tools to enhance their studies~\cite{zhu2024embracing}. The tools they use range from general-purpose assistants like ChatGPT~\cite{openai_chatgpt} to specialized educational LLMs~\cite{carnegie2024livehint,khanacademy2023khanmigo,google2024learnlm,al2024can}. By answering student questions through natural language interactions, these systems generate dialogues between students and tutors that could be a valuable resource for evaluating student understanding~\cite{scarlatos2025exploring}. In this idea paper, we propose QueryQuilt, an LLM-based system that analyzes student-LLM chat logs to identify common knowledge gaps without requiring direct instructor intervention, aiming to answer the following research question: \textbf{How can we automatically evaluate students' knowledge gaps by analyzing student-LLM chat logs?}

QueryQuilt takes a detailed knowledge component (KC) list as input, which corresponds to the course curriculum. The system comprises two LLM-based components: (1) a frontend dialogue agent that interacts with students to address their questions and lead them to expose knowledge gaps, and (2) a backend knowledge gap identification agent that processes conversations between students and the dialogue agent to identify knowledge gaps related to specific KCs from the predefined list. These identified gaps are then aggregated by frequency to provide instructors with a comprehensive view of class-wide learning challenges, enabling them to adapt their teaching plans and strategies accordingly.

Our preliminary experiments demonstrate that QueryQuilt can effectively identify both individual and collective knowledge gaps among students, indicating promising directions for future implementation and evaluation in real-world educational settings.

\textbf{Paper Structure}: Section~\ref{sec:motivation} motivates the paper through a real world dialogue example; Section~\ref{sec-related} presents the and related work; Section~\ref{sec-design} describes the QueryQuilt system architecture; Section~\ref{sec-evaluation} details our preliminary evaluation results; Section~\ref{sec-discussion} discusses limitations and future directions; Section~\ref{sec:plan-for-eval} outlines our comprehensive evaluation strategy; and Section~\ref{sec:conclusion} concludes the paper.

\section{Motivation Example}
\label{sec:motivation}
In this section, we demonstrate the richness of information that can be extracted from student-AI interactions to reveal knowledge gaps. The following dialogue comes from StudyChat~\cite{mcnichols2025studychat}, a dataset of authentic student conversations with an LLM-powered tutoring chatbot during a university-level AI course. The response from LLM are summarized for brevity.

\begin{simplebox}
\textbf{student:} what does the score for the logistic regression model measure in use .score in sklearn? \\
\textbf{AI:} [AI explains that the score in sklearn's logistic regression refers to model accuracy, measured as the proportion of correctly classified samples in the dataset] \\
\textbf{student:} do you mean that the model was 93.33\% probable to classified the data points corrcetly? \\
\textbf{AI:} [AI clarifies that 93.33\% means the model correctly classified this percentage of the test data points, while noting other metrics should also be considered] \\
\textbf{student:} In this case, the model is certain that it predicted the correct class with 93.33\% probability? \\
\textbf{AI:} [AI corrects the misconception by explaining that 93.33\% represents the confidence level for an individual prediction belonging to the positive class, not the model's overall certainty of being correct]
\end{simplebox}
\vspace{2pt}
For readers unfamiliar with machine learning concepts, logistic regression is a model that classifies data into categories by assigning probability values between 0 and 1. Values above 0.5 typically indicate one class, while values below indicate another. The \verb|.score()| method mentioned is a scikit-learn function that measures overall accuracy, the percentage of correctly classified samples.

This simple dialogue reveals complex knowledge gaps. At the most immediate level, we can see the student confuses the \verb|.score()| method's output with prediction probability. This confusion persists despite the AI's clarification,evidenced by the student's third question that continues to conflate these distinct concepts. Examing more deeply, this misunderstanding indicates that the student may confuses model-wide evaluation metrics (accuracy across a dataset) and instance-level prediction confidence (probability for a single classification). This error suggests an incomplete understanding of the machine learning evaluation framework, where different metrics serve distinct purposes at different levels of analysis.This example shown the depth of education diagnostic information could be extracted from a brief three-turn exchange, demonstrated student-AI chat logs as rich source for uncovering knowledge gaps. 

More importantly, modern LLMs demonstrate the capability to directly automate such dialogue analysis. The following text shown the analysis on the dialogue provided by Claude Sonnet 3.5~\cite{anthropic2024claude} effectively captured the knowledge gaps.

\begin{simplebox}
Looking at the conversation, I can identify the following misconceptions and knowledge gaps about logistic regression:

The student confuses the .score() method (which returns accuracy as a percentage of correctly classified instances) with the probability output for individual predictions.
There's confusion between model performance metrics (like accuracy) and prediction confidence (probability values) 
The student doesn't appear to grasp that logistic regression outputs probabilities for each instance individually, not a single probability value for the entire model's predictive capability.
\end{simplebox}
\vspace{2pt}
Based on this insight, We propose QueryQuilt, a novel LLM-based framework for automated knowledge gap detection in lectures, designed to provide instructors with timely evaluation on student understanding.

\section{Related Work}
\label{sec-related}
\subsection{Student Response Systems}
Student Response Systems (SRS), also known as classroom communication systems~\cite{white2011clicking}, audience response systems~\cite{connor2011using, cain2008primer}, or voting systems~\cite{king2016investigating, mathiasen2015digital}, are interactive technologies that help instructors create engagement through various question formats, including multiple-choice and short-answer questions. Research has shown that proper implementation of such systems can significantly increase student engagement~\cite{shon2011review, kappers2015poll} and help instructors gauge student understanding of course material~\cite{pai2025using}. However, SRS effectiveness is limited by its instructor-driven nature: system performance depends on instructor technology familiarity and question design~\cite{nielsen2013teaching}, and can only capture student understanding during active instructor-initiated interactions, missing spontaneous confusions. Moreover, studies report that improperly used SRS could even harm the learning output~\cite{wood2020systematic}. Participation bias (15-20\% of students resist collaborative environments~\cite{hoekstra2015because}) also limits comprehensiveness of the evaluation\cite{wood2020systematic,nielsen2013teaching}. Overall, SRS remains fundamentally reactive rather than proactive in identifying knowledge gaps.

\subsection{Automated Knowledge Gap Detection}
Researchers have proposed various systems to identify student knowledge gaps. Traditional approaches include statistical methods \cite{lan2015mathematical}, NLP-based systems using sentence embeddings \cite{botelho2023leveraging}, and rule-based feedback systems \cite{kochmar2020automated}, these methods requiring significant human effort to adapt to new domains due to limited model capacity. Recent LLM-based advances have shown improved capabilities in both gap identification and feedback generation, including examining code behavior explanations \cite{oli2024can}, identifying false mathematical reasoning patterns \cite{mcnichols2023automated}, and providing adaptive tutoring \cite{wang2025training} and debugging assistance \cite{al2023socratic}. Another related field is Knowledge Component (KC) assignment, where researchers have developed tools to assign knowledge components to question responses for learning evaluation \cite{choffin2019das3h, pavlik2009performance}. Recent advances include using LLMs to automatically discover high-quality KCs from mathematics questions \cite{moon2024using}. LLMKT \cite{scarlatos2025exploring} proposed a method for identifying KCs from dialogues between students and expert formative assessment tutors. Despite these advances, most methods still work with structured problem sets rather than natural student conversational history. While LLMKT operates with dialogue data, it focuses on students' interactions with tutors using specifically designed questions rather than general dialogue with the LLM, thus limiting applicability to more natural classroom interactions.
\subsection{Dialogue Analysis}
Existing works on dialogue analysis mainly focus on analyzing tutors' behaviors in dialogue to evaluate the effectiveness of tutoring, such as evaluating the effectiveness of tutors' "Talk Acts" \cite{abdelshiheed2024aligning, vail2014identifying, boyer2008learner}. There are a few works that directly operate on dialogue data: Ikram et al. \cite{ikram2025exploring} use LLMs to predict future tutor behaviors using existing dialogue data, and LLMKT \cite{scarlatos2025exploring} directly analyzes students' knowledge gaps through examining their dialogue with tutors.

\section{QueryQuilt}
\label{sec-design}

QueryQuilt is designed to achieve two primary objectives: First, it functions as a regular AI assistant answering student questions while simultaneously identifying student knowledge gaps. Second, it effectively aggregates these identified gaps to generate a comprehensive class-level overview, providing instructors with valuable reference data for intervention. For the first objective, QueryQuilt adopts a two-agent framework with each agent responsible for dialogue and knowledge gap detection respectively, effectively decoupling regular assistance from knowledge gap identification. For the second Objective, QueryQuilt uses a KC list as a standardized intermediate layer to represent knowledge gaps, enabling straightforward knowledge gap aggregation.

\begin{figure} 
\centering \includegraphics[width=0.45\textwidth]{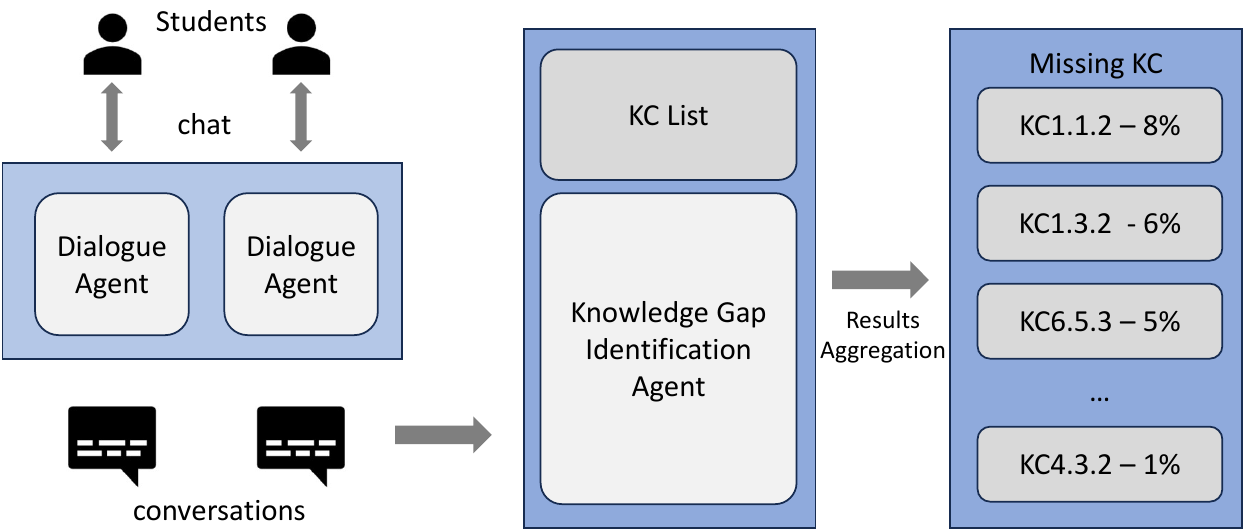} 
\caption{Overview of QueryQuilt Workflow}
\label{fig:system-architecture}
\end{figure}

Figure~\ref{fig:system-architecture} illustrates the overall system architecture and workflow. QueryQuilt employs two specialized LLM agents: A \textit{Dialogue Agent} that interacts with students and prompted to guide students to reveal their knowledge gaps, and a \textit{Knowledge Gap (KG) Identification Agent} that analyzes student chat logs and maps identified knowledge gaps to standard knowledge components provided in the KC list. The system then counts the frequency of each KC to identify common knowledge gaps across the classroom. 

\subsection{Dialogue Agent}
The Dialogue Agent serves as the primary student-facing interface, functioning as an intelligent tutoring AI assistant that provides real-time question answering while strategically leading students to expose their knowledge gaps. The dialogue agent uses retrieval-augmented generation (RAG)~\cite{lewis2020retrieval} to generate content aligned with the course material by retrieving relevant information before formulating responses. Unlike traditional AI assistants that focus solely on providing answers, the dialogue agent is explicitly prompted to include probing questions at the end of responses to naturally lead students to provide more information about their thinking process. The following is part of the system prompt of the dialogue agent, demonstrating how we instruct it to balance direct assistance with pedagogical inquiry to enhance learning outcomes while guiding students to express their understanding.
\begin{simplebox}
\textbf{Interaction Strategy}\\
1. First, acknowledge their specific issue and offer a solution \\
2. As you explain the solution, naturally incorporate questions that help reveal their conceptual understanding\\
3. Conclude with follow-up questions as next steps\\
\textbf{Examples of Natural Diagnostic Questions:}\\
1. It seems you might be encountering an issue with [concept]. How do you typically think about [concept] when you're designing your solution?\\
2. What's your goal after this step? Knowing that will help me suggest the most appropriate method.
\end{simplebox}

\subsection{Knowledge Gap Identification Agent}
The KG identification Agent serves as the core analytical component that processes student dialogues to identify knowledge gaps. To facilitate the later knowledge gap merge, instead of asking the LLM to directly point out the knowledge gap from chat logs, this agent maps student knowledge gap to a detailed KC lists includes fine grained knowledge components covered in the course. The instructor is responsible to providing the KC lists for analysis. 

KG identification agent adopted KC lists in a similar format of Common Core Standard of Math~\cite{CCSSI2010math}. Knowledge Component~\ref{lst:kclist} shows an example knowledge component for the KC list. knowledge components are managed with a hiearchy from broad general concept to specific knowledge component and students behavior. The entire KC list is included in the system prompt of KG identification agent as a reference. 

\begin{graybox}{1. Machine Learning Fundamentals
}{ML-F}
\caption{A knowledge component for KC list.}
\label{lst:kclist}
\textbf{2. Distinguish between supervised and unsupervised learning methods and their appropriate applications}
\begin{enumerate}
  \item Explain the fundamental difference between supervised and unsupervised learning, and how to select the appropriate method based on data characteristics, objectives, and available resources. For example, we choose supervised learning algorithms when we have labeled datasets with input features X and target variables Y because we want the model to learn the mapping function f(X)=Y; whereas when we only have input data X without corresponding output labels, we use unsupervised learning to discover hidden patterns or structures within the data.
\end{enumerate}
\end{graybox}

For each dialogue turn pair $(t_j, s_j)$, where $t_j$ represents the dialogue agent turn and $s_j$ the student response, the system evaluates the student's answer for correctness and identifies specific knowledge gaps. QueryQuilt records the identified KCs and associated misconceptions from all students, maintaining a frequency table that tracks the prevalence of each knowledge gap across the student population. This aggregated data is then output for the instructor as a reference.

\section{Preliminary Evaluation}
\label{sec-evaluation}
We evaluated the feasibility and effectiveness of QueryQuilt though two different experiments: (1) an end-to-end tests using LLM generated students to evaluate the feasibility of all components work as a whole. (2) An KC-extraction experiment using real world students conversation data to provides more realistic evaluation. We conducted our experiments using Claude 3.5 Sonnet, an advanced large language model released by Anthropic~\cite{anthropic2024claude} and a KC list contains 296 unique KCs for an undergraduate artificial intelligence(AI) course developed by an ex-instructor .

\subsection{Simulated Study with Generative Students}
To evaluate the end-to-end effectiveness of our system in a controlled environment, we test QueryQuilt with LLM simulated students exhibiting specific knowledge gaps.  We utilized prompt engineering techniques of generative student~\cite{lu2024generative} to ensure realistic student-system interactions. We constructed a scenario involving an undergraduate AI course and generated 20 distinct simulated student profiles, organized into four groups of five students each. Students within each group shared a common missing knowledge component, while individually manifesting different behavioral patterns in relation to their knowledge gap. We evaluated system performance using four quantitative metrics:

(1) \textbf{Detection rate}: The proportion of conversations in which the KC-extraction agent successfully identified the predefined missing knowledge component in its output list.

(2) \textbf{Speed of detection}: The conversation turn at which the KG identification agent first correctly identified the missing knowledge component. This metric provides insight into the dialogue agent's efficiency in exposing student knowledge gaps.

(3) \textbf{Top-1 accuracy}: The percentage of conversations where the highest-confidence knowledge component identified by the KG identification agent matched the actual missing KC defined in the student profile, serving as a precision measure for our dialogue agent's elicitation capability.

(4) \textbf{KC frequency distribution}: The aggregated count of different knowledge components identified across all simulated students, demonstrating QueryQuilt's ability to detect common knowledge gaps across a student population.

Table~\ref{tab:agent-comparison} depicted the performance of QueryQuilt on identifying knowledge gap on individual level, QueryQuilt detection rate of 100\%, successfully identifying all predefined knowledge gaps across the simulated student population within 1.3 turns of dialogue. and Top-1 accuracy 80.0\%.

\begin{table}[t]
\centering
\caption{Naive vs. Strategic Agent Performance}
\label{tab:agent-comparison}
\begin{tabular}{lccc}
\toprule
\textbf{Agent} & \textbf{Det.(\%)} & \textbf{Speed (\# of turns)} & \textbf{Top-1(\%)} \\
\midrule
Probing & 100 & 1.3 & 80.0 \\
\bottomrule
\end{tabular}

\end{table}

Figure~\ref{fig:top-10} illustrates the frequency distribution of identified knowledge components across simulated student interactions. QueryQuilt successfully identified all four ground truth knowledge components (KC1.6.1, KC2.4.1, KC1.3.1, and KC1.2.1) as their most frequently detected KCs, demonstrating the system's robust ability to aggregate common knowledge gaps across multiple students. This finding validates QueryQuilt's core functionality: the automatic detection of widespread knowledge gap without instructor intervention.

\begin{figure} 
\centering 
\includegraphics[width=0.45\textwidth]{imgs/dialogue-kc-top10.pdf} 
\caption{Frequency distribution of identified knowledge gaps in the simulated student experiment. Ground truth knowledge gaps are indicated with darker bars and asterisk. Only the top 10 most frequently identified components are shown for clarity.}
\label{fig:top-10}
\end{figure}

Overall, this experiment validates QueryQuilt's capability on identifying and aggregate knowledge gaps across students' chat log. However, as simulated students has inherent limitations in representing the complexity and diversity of real educational interactions. A comprehensive evaluation with real students and class context of the system's performance in practical settings is needed.

\subsection{Knowledge Gap Extraction on Real Data}
To evaluate the QueryQuilt's capacity of identifying Knowledge Gap on real world data. In this experiment, We applied QueryQuilt to StudyChat~\cite{mcnichols2025studychat}, a dataset with real-world student conversations with an LLM-powered tutoring chatbot in a semester-long, university-level artificial intelligence (AI) course. We sampled 20 dialogues from the dataset where the dialogues' topics closely related to the course material, and manually labeled each of them with knowledge gap identified in the form of knowledge components, for each dialogue, we identified 1-3 knowledge components as the knowledge gap and We send these 20 dialogues to QueryQuilt and compare the identified knowledge gap against the human labeling. 

We consider the a success if the knowledge gaps identified by the QueryQuilt includes the knowledge gaps we labeled (could have more but not less). Under this setting. The output of QueryQuilt successfully identified all human labeled knowledge components for 19 of the 20 dialogues. The only dialogue that where QueryQuilt failed to identify the knowledge gap is because the LLM think the dialogue is too short and cannot be used to derive reliable conclusion from it. 
This experiment demonstrated the capacity of LLM on extracting knowledge gaps from real world diagnosis.

\section{Limitations and Discussion}
\label{sec-discussion}
Based on our preliminary results, QueryQuilt has demonstrated significant potential for utilizing chat histories to analyze students' knowledge gaps. However, several challenges emerged during our experiments that warrant consideration for future iterations of QueryQuilt, particularly when implementing the system in actual classroom environments.

\subsection{Student Behavior}
Although our experiments indicated that QueryQuilt could identify collective misunderstandings among LLM-simulated students, we observed substantial differences between the behavior of simulated students and real students. Simulated students frequently volunteered their misconceptions even when explicitly prompted not to do so. For instance, in one conversation, despite QueryQuilt providing a comprehensive solution, the simulated student still posed a question revealing a knowledge gap: \textit{``I totally understood! Just to make sure - once I get good accuracy on the validation set, that means my model is ready for submission, right?''}

Furthermore, simulated students consistently demonstrated a willingness to share background information systematically and responded to questions posed by the assistant, which facilitated easier detection of knowledge gaps. In contrast, real students exhibited more diverse interaction patterns. A significant proportion used very brief prompts such as \textit{``implement rnn using torch''} or simply copied content from textbooks and asked the AI assistant for explanations without follow-up questions, thereby limiting the information available in the dialogue.

These behavioral differences present challenges for deploying dialogue agents in real-world educational settings. Designing an effective dialogue agent requires balancing the need for probing questions that asking additional information from students against the providing comprehensive information typical of standard AI assistants. Providing excessive information may result in students using short prompts, which influence the ability to identify knowledge gaps. Conversely, adding too many probing questions without offering concrete solutions may reduce students' motivation to use the system, potentially causing them to revert to other platforms.

\subsection{Knowledge Component Lists}
A critical observation from our experiments is that accurate identification of knowledge gaps through the KG identification agent requires high-quality knowledge component lists. These lists must be sufficiently detailed to enable the agent to identify fine-grained knowledge gaps, while also being comprehensive enough to cover most potential questions that may arise during the course.

Researchers in K-12 education frequently utilize the Common Core Standards~\cite{CCSSI2010math}, which define fine-grained knowledge components. However, such comprehensive KC lists are often lacking for college-level education. Finding ways to facilitate the generation of such KC lists also represents a valuable direction for future research.

\section{Plan for Further Evaluation}
\label{sec:plan-for-eval}
we will develop a web service to deploy QueryQuilt in an undergraduate data structures course at University A in Spring 2026. The system will be accessible to students via browser during lectures. Our evaluation will focus on two primary aspects: validating knowledge gap detection accuracy and assessing stakeholder experience with the system.

\subsection{Knowledge Gap Detection Validation}
We will assess QueryQuilt's accuracy through a real-time feedback cycle during each lecture. The system will process student-AI dialogues to identify prevalent knowledge gaps, providing instructors with a ranked list of the top 3-5 issues 15 minutes before class ends. Instructors will then rapidly develop targeted quiz questions addressing these specific gaps, which students will complete in a brief 5 minute end-of-class assessment. By analyzing the correlation between system-identified knowledge gaps and student performance on these targeted questions, we can directly validate the system's detection capabilities and measure the impact of just-in-time instructional interventions.

\subsection{System Utility and Stakeholder Experience}
We will collect qualitative feedback through instructor interviews (n=2) and student surveys (n=100) to assess the system's utility and perceived learning value. The interviews and surveys will specifically measure whether instructors find the real-time knowledge gap detection actionable for immediate instructional adjustments and whether students perceive value in the dialogue interactions. Additionally, we will monitor student participation metrics, including usage frequency, conversation depth, and overall adoption rate during lectures. This engagement data will provide critical insights into student willingness to interact with AI-based educational tools and help identify factors that influence sustained usage. The combined quantitative and qualitative measures will offer a comprehensive view of QueryQuilt's integration into the learning environment and its acceptance by key stakeholders.

\section{Conclusion}
\label{sec:conclusion}
QueryQuilt presents a novel multi-agent LLM framework that automatically identifies student knowledge gaps by analyzing AI-student dialogues. Our preliminary evaluations with both simulated and real student data demonstrate promising capabilities in detecting and aggregating knowledge gaps automatically. We identified potential challenges including differences in simulated and real student interaction patterns and the need for comprehensive knowledge component lists. As future work we plan to deploy QueryQuilt in an authentic undergraduate data structures course to further validate the approach's effectiveness.


\clearpage
\bibliographystyle{ACM-Reference-Format}
\bibliography{main}

\appendix

\end{document}